\documentclass[12pt]{article}
\makeatletter
\def\eqnarray{%
 \stepcounter{equation}%
 \let\@currentlabel=\theequation
 \global\@eqnswtrue
 \global\@eqcnt\z@
 \tabskip\@centering
 \let\\=\@eqncr
 $$\halign to \displaywidth\bgroup\@eqnsel\hskip\@centering
 $\displaystyle\tabskip\z@{##}$&\global\@eqcnt\@ne
 \hfil$\displaystyle{{}##{}}$\hfil
 &\global\@eqcnt\tw@$\displaystyle\tabskip\z@{##}$\hfil
 \tabskip\@centering&\llap{##}\tabskip\z@\cr}
\makeatother

\renewcommand{\theequation}{\thesection.\arabic{equation}}

\usepackage{amssymb}

\date{\empty}

\begin{document}

\title{$N=(4,0)$ Super-Liouville Theory \\ on the Coadjoint Orbit and PSU(1,1$|$2)}
  
\author{Shogo Aoyama\thanks{Professor emeritus, e-mail: aoyama.shogo@shizuoka.ac.jp}
 \\
       Department of Physics,  
              Shizuoka University \\
                Ohya 836, Shizuoka  \\
                 Japan}

\maketitle

\vspace{2cm}

\begin{abstract}
An $N=(4,0)$ supersymmetric Liouville theory is formulated by the coadjoint orbit method. It is discovered that the action has  symmetry under PSU(1,1$|$2).
\end{abstract}

\vspace{5cm}

\noindent
Keywords

2D effective gravity, Extended supersymmetry, Differential geometry, Superconformal algebra

\newpage

\section{Introduction}

In the past few years there has been much interest in a duality between the SYK model and the  ${\rm D}=2$ effective gravity. The Schwarzian  theory is considered as playing a role of  a mediator between the two. The differential geometrical aspects of the Schwarzian theory  got clarified when it was reformulated by the coadjoint orbit method in \cite{Wi}.
 Supersymmetric extensions of the theory have been also discussed\cite{Super}, but without 
 a proper account on the differential geometry.  
Recently the extension to the  $N=4$ case   has been undertaken by means of the coadjoint orbit method in \cite{AH2} by following the work  \cite{Wi}. The  obtained  theory has  been shown to have symmetry under  PSU(1,1$|$2). The arguments there can be straightforwardly applied for the lower symmetric Schwarzian theories.

 The aim of this letter is 
to formulate the $N=(4,0)$ super-Liouville theory in 
1+1 dimensions by similarly working out the coadjoint orbit method. The Liouville  theory is the simplest one for the ${\rm D} =2$ effective 
gravity.  
 After  Polyakov's work it was extensively studied by  various methods. (See \cite{A1, Se} for an overview of the studies at the early stage.) Among them the coadjoint orbit method, which was  proposed by Alekseev and Shatashvili\cite{Al}, is the most geometrical. 
After this work the coadjoint orbit method  was generalized to get the $(1,0)$ and $(2,0)$ supersymmetric Liouville  theories in \cite{A2} and \cite{Nieu} respectively.  The left-moving sectors of the respective theories were  extended so as to admit the $N=1$ and $N=2$ superconformal symmetries. 
On the other hand in  the right-moving sectors the conformal symmetry remained  non-supersymmetric, but  the  symmetry SL(2) got promoted to OSp(2$|$1) and  OSp(2$|$2) for the respective   theories\cite{A3,A4}. A further extension of the coadjoint orbit method  to  the $N=(4,0)$ case  has not been discussed.

The letter is organized as follows. In Section 2 we give  a short summary of the $N=(4,0)$ superconformal diffeomorphism and the $N=4$ super-Schwarzian derivative, following \cite{AH2}. In Section 3 we then work out 
the coadjoint orbit method to get  the $N=(4,0)$ super-Liouville theory as (\ref{non-local action}).  In contrast with the lower symmetric theories it is given in a non-local form like the WZWN model. The author believes that there is no way to find a local expression 
 as long as we stick to the supercovariant formalism using the superfields. 
A local expression of the action is found as (\ref{actionlocal}) by expanding it 
in components.  We then check that 
 the purely bosonic part of the action coincides with  the known non-supersymmetric  Liouville theory. After this check we  calculate  
the energy-momentum tensor of the theory by using the action of the non-local form  (\ref{non-local action}). It is found to be given by the $N=(4,0)$ super-Schwarzian derivative. In Section 4  we show that 
action (\ref{non-local action}) has  symmetry under PSU(1,1$|$2). It is realized on the coset space PSU(1,1$|$2)/\{SU(2)$\times$U(1)\} of which holomorphic coordinates are superfields $f,\varphi_a,\varphi^a$ representing the $N=(4,0)$ superconformal diffeomorphism.   
The same symmetry has been discovered  for the $N=4$ super-Schwarzian theory in \cite{AH2}.

Appendix A is devoted to a summary on  the supersymmetric extension of the Liouville theory.  
 In Appendix B we prove some formulae  which are assumed in the main body of this letter. At important steps of the arguments 
 we have recourse to expansion of $N=(4,0)$ superfields in components, as in \cite{AH2}. But the calculations are  much more complicated here. The details of them are presented in a separate note \cite{note}.

\section{The (4,0) superconformal diffeomorphism}
\setcounter{equation}{0}

The $N=(4,0)$ superconformal group is a group of which elements are superdiffeomorphisms in the 
 $N=(4,0)$ superspace.    The $N=(4,0)$ superspace is described by the supercoordinates
\begin{eqnarray}
  (x, \theta_1,\theta_2,\theta^1,\theta^2  ,t) \equiv (x, \theta ,t). \label{coord}
\end{eqnarray}
Here $x$ is a real coordinate. $\theta_a,a=1,2$, are fermionic ones, while $\theta^a,a=1,2$, their complex conjugates.  They consist of the coordinates of the supersymmetric sector, while $t$  a real coordinate of the non-supersymmetric sector. The respective sectors are called the left- and right- moving sectors.  
 The details of the  superconformal diffeomorphism  in the left-moving  has been discussed in
 \cite{AH2}. 
Namely we consider superdiffeomorphism in the left-moving sector
\begin{eqnarray}
x &\longrightarrow&  f(x, \theta,t), \quad 
\theta_a \longrightarrow\varphi_a(x, \theta,t),  \quad
\theta^a \longrightarrow\varphi^a(x, \theta,t). \quad\quad  \label{finitediffeo}
\end{eqnarray}
Its infinitesimal form is given by 
\begin{eqnarray}
x &\rightarrow& x+ \delta_v f|_{(f,\varphi)=(x,\theta)}, \ \
\theta_a \rightarrow \theta_a+\delta_v \varphi_a|_{(f,\varphi)=(x,\theta)},  \ \
\theta^a \rightarrow \theta^a+\delta_v\varphi^a|_{(f,\varphi)=(x,\theta)}. \nonumber
\end{eqnarray}
A superconformal field with weight $(w,0)$ is defined as transforming by these diffeomorphisms as
\begin{eqnarray}
\delta_v \Psi_{(w,0)}=  [v\partial_x+{1\over 2}D_{\theta c}v D^{\ c}_\theta+{1\over 2} D^{\ c}_\theta\xi D_{\theta c}+w\partial_xv ] \Psi_{(w,0)}, \label{h,0}
\end{eqnarray} 
with 
$$
v=\delta f|_{(f,\varphi)=(x,\theta)}+\theta_c\delta\varphi^c|_{(f,\varphi)=(x,\theta)}+ \theta^c\delta\varphi_c|_{(f,\varphi)=(x,\theta)}.
$$
The superfields $f,\varphi_a,\varphi^a$ describing  the superconformal diffeomorphisms (\ref{finitediffeo}) may be  given  by superfields with weight (0,0).  They satisfy the superconformal conditions. Moreover $\varphi_a$ and $\varphi^a$     
 satisfy  the chirality conditions.   
  That is,   
\begin{eqnarray}
\delta_v f&=&[v\partial_x+{1\over 2}D_{\theta c} v D_{\theta}^{\ c}+ 
{1\over 2}D_{\theta }^{\ c}v D_{\theta c}]f, \label{superconf1} \\
\delta_v \varphi_a&=&[v \partial_x+{1\over 2}D_{\theta c}v D_\theta^{\ c}]\varphi_a,  \label{superconf2}\\
\delta_v \varphi^a&=&[v \partial_x
 +{1\over 2}D_{\theta }^{\ c}v D_{\theta c}]\varphi^a. \label{superconf3}
\end{eqnarray}
As in \cite{AH2} the super-Schwarzian derivative for the $N=(4,0)$ case  is given by 
\begin{eqnarray}
{\cal S}(f,\varphi;x,\theta,t)=\log\det[D_{\theta a}\varphi^b].  \label{S}
\end{eqnarray} 
By the superdiffeomorphisms (\ref{superconf1})$\sim$(\ref{superconf3})
it transforms anomalously 
\begin{eqnarray}
\delta_v{\cal S}(f,\varphi;x,\theta,t)=[v\partial_x+{1\over 2}D_{\theta c} v D_{\theta}^{\ c}+ {1\over 2}D_{\theta }^{\ c}v D_{\theta c}]{\cal S}(f,\varphi;x,\theta)+\partial_x v.  
 \label{superdiff}
\end{eqnarray}

\section{ The $(4,0)$ super Liouville theory}
\setcounter{equation}{0}

Now we construct an $N=4$  super-Liouville theory   by applying  
 the coadjoint orbit method for the  
superconformal group so far reviewed. 
The  superconformal algebra $\frak g$  and  the dual space  ${\frak g}^*$ are centrally extended. Their elements  are given by 
\begin{eqnarray}
(u ,k)\in {\frak g}, \quad\quad (b ,c)\in {\frak g}^*.  \nonumber
\end{eqnarray} 
Here $k$ and $c$ are central elements. 
$u$ and $b$ are bosonic superfields,  obeying the superconformal transformations of $\Psi_{(-1,0)}$ and $\Psi_{(0,0)}$ given by  (\ref{h,0}) respectively. But the latter transformation becomes possibly
 anomalous. 
 The volume element $dxd^4\theta$ for  the left-moving sector  of the $N=(4,0)$ superspace  has weight 1,
 so that 
   the invariant quadratic form is defined  by 
\begin{eqnarray}
<(b,c),(u,k)>=\int dxd^4\theta\ b u+ck.   \label{invform}
\end{eqnarray}
The  centrally extended superconformal algebra $\frak g$ is given  by  the infinitesimal adjoint action ${\rm ad}_{(v,l)}$ on $(u,k)\in {\frak g}$
\begin{eqnarray}
{\rm ad}(v,l)(u,k) 
&=& \Big( v\partial_x u-u\partial_x v +{1\over 2}D_{\theta c} v D_\theta^{\ c} u +{1\over 2}D^{\ c}_\theta v D_{\theta c} u , \int dxd^4\theta\ v\partial_x u\Big) \nonumber \\
&\equiv& [(u,k),(v,l)].    \label{commut}
\end{eqnarray}
 Using the relation 
\begin{eqnarray}
<{\rm ad}^*(v,l)(b,c),(u,k)>=-<(b,c),{\rm ad}(v,l)(u,k)>,   \nonumber
\end{eqnarray}
we then find 
  the corresponding coadjoint action ${\rm ad}^*(v,l)$ on $(b ,c)\in {\rm g}^*$  
\begin{eqnarray}
{\rm ad}^*(v,l)(b,c)=\Big([v\partial_x+{1\over 2}D_{\theta c}v D_\theta^c+{1\over 2} D_\theta^c v D_{\theta c}]b+c\partial_x v,0\Big), \label{coadjoint}
\end{eqnarray} 
 which is also centrally extended. 
  We think of a coadjoint orbit O$_{(b,c)}$, whose initial point is $(b,c)\in {\rm g}^*$. 
The finite form of (\ref{coadjoint}) is generated on the coadjoint orbit  by the superdiffeomorphism (\ref{finitediffeo})
 as\footnote{ Here the arguments of $b$ have been explicitly written as $b(f,\varphi,t)$.  
 As in \cite{AH2} our convention is that superfields always depend on $(x,\theta,t)$ if any argument is not written.}
\begin{eqnarray}
{\rm Ad}^*(f,\varphi)(b,c)&\equiv&\Big(b(f,\varphi,t)+c{\cal S}(f,\varphi;x,\theta,t), c\Big).  \label{Ad*}
\end{eqnarray}
Here ${\cal S}(f,\varphi;x,\theta,t)$ is the super-Schwarzian derivative given by (\ref{S}). 
Now we consider the right-moving sector 
in an enlarged  space  with coordinates  $(t_1,t_2,\cdots,t_n)$, called ${\cal O}_n$. 
 That is, the supercoordinates (\ref{coord}) describing the $N=(4,0)$ superspace  are extended as
\begin{eqnarray}
(x, \theta_1,\theta_2,\theta^1,\theta^2  ,t_1,t_2,\cdots,t_n) \equiv (x, \theta ,t).
\end{eqnarray} 
The Kirillov-Kostant 2-form  is  given   by 
\begin{eqnarray}
\Omega_{(b,c)}={1\over 2}<{\rm Ad}^*(f,\varphi)(b,c),[(y,0),(y,0)]>, 
 \label{KK2}
\end{eqnarray}
on the coadjoint orbit $O_{(b,c)}$ in ${\cal O}_n$.  
Here $y$ is a  
${\frak g}$-valued 1-form in ${\cal O}_n$, while $f$ and $\varphi$ are 0-forms. 
  They are superfields with the coordinates $x, \theta,t_1,t_2,\cdots,t_n$.
It is determined so that 
 the exterior derivative of the quantity (\ref{Ad*}), which is an element of ${\frak g}^*$, is induced 
by the infinitesimal coadjoint action (\ref{coadjoint}) on it along  the orbit O$_{(b,c)}$  as
\begin{eqnarray}
d{\rm Ad}^*(f,\varphi)(b,c)={\rm ad}^*(y,0)\Big(b(f,\varphi)+c{\cal S}(f,\varphi;x,\theta),c\Big). \label{dY}
\end{eqnarray}
Keep in mind that the exterior derivative acts only on the coordinates $t_1,t_2,\cdots$.  
 In \cite{AH2} $y$ is found to be a solution to this equation  such that 
\begin{eqnarray}
y={1\over \Delta}(df+ \varphi_c d\varphi^c+\varphi^cd \varphi_c),
 \label{Y}
\end{eqnarray} 
with 
\begin{eqnarray}
\Delta=\partial_x f+\varphi_c\partial_x\varphi^c+\varphi^c\partial_x\varphi_c
=\det [D_{\theta a}\varphi^{b}] \label{Delta}.
\end{eqnarray}
The centrally extended  commutator in (\ref{KK2}) becomes  
\begin{eqnarray}
[(y,0),(y,0)] =\Big(2y\partial_x y+D_{\theta c} y D_\theta^{\ c} y, \int dxd^4\theta\ y\partial_x y \Big),   \label{yy}
\end{eqnarray}
from (\ref{commut}). 
By the definition $d\Omega_{(b,c)}=0$ so that it may be locally expressed 
such that 
\begin{eqnarray}
\Omega_{(b,c)} = d\alpha.   \label{alpha}
\end{eqnarray}
Integrating this 1-form on the orbit  $ O_{(b,c)}$ gives an $N=(4,0)$ supersymmetric action   
\begin{eqnarray}
I=\int_{O_{(b,c)}} \alpha.  \label{action}
\end{eqnarray}
We propose that this is the (4,0) super-Liouville theory.

We put the Kirillov-Kostant 2-form (\ref{KK2}) in an explicit form  as
\begin{eqnarray}
2\Omega_{(b,c)}=\int dxd^4\theta \Big[\Big(b(f,\varphi,t)+c{\cal S}(f,\varphi;x,\theta,t)\Big)(2y\partial_x y+D_{\theta c} y D^{\ c}_\theta y)
+cy\partial_x y\Big],  \nonumber 
\end{eqnarray}
by (\ref{invform}) with (\ref{Ad*}) and (\ref{yy}). 
 Choose  $b$  to be zero at the initial point of the orbit for simplicity. Then by integration  by part it becomes
\begin{eqnarray}
2\Omega_{(0,c)}=c\int dxd^4\theta \Big[{\cal S}(f,\varphi;x,\theta,t)(2y\partial_x y+D_{\theta c} y D^{\ c}_\theta y)
+y\partial_x y\Big]. \label{Omega'}  
\end{eqnarray} 
To proceed with our argument we need  the following formulae
\begin{eqnarray}
d{\cal S}(f,\varphi;x,\theta,t) &=& [y\partial_x+{1\over 2}D_{\theta c}y D_\theta^c+{1\over 2} D_\theta^c y D_{\theta c}]{\cal S}(f,\phi;x,\theta,t)+\partial_x y,  \label{dS}\\
d y&=& y\partial_x y+{1\over 2}D_{\theta c} y D^{\ c}_\theta y.  \label{dy}
\end{eqnarray}
They are shown in Appendix B. 
By means of these formulae 
 the Kirillov-Kostant 2-form (\ref{Omega'}) becomes
 \begin{eqnarray}
2\Omega_{(0,c)}=c\int dxd^4\theta \Big[-d\Big(2y{\cal S}(f,\varphi;x,\theta,t)\Big)-y\partial_x y   \Big]. 
\label{KKexplicit}  
\end{eqnarray}
$\Omega_{(0,c)}$ is closed so that\footnote{(\ref{Omegaclosed}) is obvious by the definition of
 $\Omega_{(0,c)}$, but  it may  be directly checked by (\ref{dy}).}
\begin{eqnarray}
   d\int dxd^4\theta\ y \partial_x y=0.  \label{Omegaclosed}
\end{eqnarray}
 Then the anomaly term takes an exact form   such that 
\begin{eqnarray}
  \int dxd^4\theta\  y\partial_x y =d\gamma,  \label{gamma}
\end{eqnarray}
with  a 1-form $\gamma$. 
To study  $\gamma$  it is helpful to remember the arguments for the lower supersymmetric cases.  The  Kirillov-Kostant 2-form  takes the form  (\ref{KKexplicit}), in which all the quantities  are replaced by  those for the lower supersymmetries, given in Appendix A. In particular the anomaly term is replaced by
 \begin{eqnarray}
&\ &\hspace{1cm}  \int dx\ y\partial_x^3 y, \hspace{4.8cm}\ \ \ {\rm for}\  N=(0,0),  \nonumber\\
&\ &\hspace{1cm}\int dxd\theta\ {1\over 2}yD_\theta\partial_x^2 y,   
 \hspace{3.2cm} \hspace{0.55cm}\ \ {\rm for} \ N=(1,0), \nonumber\\
&\ & \hspace{1cm}\int dxd^2\theta\ {1\over 2}y\partial_x[D_{\theta +},D_{\theta_ -}]y,\quad\quad \hspace{1cm}\hspace{0.6cm}\ {\rm for} \ N=(2,0), \hspace{0.4cm} \nonumber
\end{eqnarray}
which  are closed as well. 
For these  we can find local expressions of $\gamma$ as 
\begin{eqnarray}
 \gamma &=& -\int dx y\Big[{\cal S}+{1\over 2}({\partial_x^2 f\over \partial_x f})^2\Big], \hspace{2.6cm} {\rm for}\ N=(0,0),  \nonumber\\
    \nonumber \\
\gamma &=&-\int dxd\theta y\Big[{\cal S}+{D_\theta^3\varphi\over D_\theta\varphi}{D_\theta^2\varphi\over D_\theta\varphi}\Big],   
   \hspace{1.2cm}\ \ \hspace{0.5cm} {\rm for}\  N=(1,0), \nonumber\\
    \nonumber \\
  \gamma&=& -\int dxd^2\theta y\Big[{\cal S}-2{\partial_x\varphi^+\over D_{\theta +}\varphi^+}{\partial_x\varphi^-\over D_{\theta -}\varphi^-}  \Big],  \quad\ \  {\rm for}\ N=(2,0). \nonumber
\end{eqnarray}
with ${\cal S}$ the Schwarzian derivative for the respective  supersymmetric case. 
 Owing to these formulae  the Kirillov-Kostant 2-form (\ref{KKexplicit}) takes the exact form (\ref{alpha}), in which the 1-form $\alpha$ is explicitly given.   
 The resulting action   (\ref{action})
 gives  the known Liouville actions  for the lower supersymmetric cases, summarized in Appendix A.
On the contrary,  we could  not find a local expression of $\gamma$ for
 (\ref{gamma}). Presumably it would not be possible at all in the supercovariant formulation with the 
superfields. Hopefully it would be possible if the anomaly term $y\partial y$ is expanded in components. To see this 
 we have done  rather massive calculations, using the expansion formulae of 
 $f,\varphi_a,\varphi^a$  discussed in Appendix A in \cite{AH2}. We have indeed found  it in an exact form as  (\ref{gamma}). The details of the calculations  were exposed in \cite{note}. 
  We quote here  only the result\footnote {Our convention is 
that $\int d^4\theta(\theta\cdot\theta)^2=1$}
\begin{eqnarray}
\gamma &=&2\int dx\Big\{{1\over 2}dh \Big[\partial_x^2(-{1\over\rho\xi})+\partial_x\log(\rho\xi) 
\partial_x(-{1\over\rho\xi})\Big] \nonumber\\  
&\ & +4{dh\over(\rho\xi)^2}\Big[\Big(\partial_x^2(\rho\eta)\cdot
\partial_x(\xi\eta)\Big) -
\Big(\partial_x(\rho\eta)\cdot
\partial_x^2(\xi\eta)\Big)\Big]    \nonumber\\
&\ & \hspace{0cm} -{2\over\rho\xi}\Big[\Big(d\partial_x(\rho\eta)\cdot\partial_x(\xi\eta)\Big)
-\Big(\partial_x(\rho\eta)\cdot d\partial_x(\xi\eta)\Big)\Big]
\nonumber\\
&\ & \hspace{0cm} -{1\over\rho\xi}\partial_x\Big[\Big(\partial_x(\rho\eta)\cdot d(\xi\eta)\Big)-
\Big(d(\rho\eta)\cdot \partial_x(\xi\eta)\Big)\Big]
\nonumber\\   
&\ & \hspace{0cm}+{1\over 2}\Big[\Big(\rho\eta\cdot d(\xi\eta)\Big)-\Big(d(\rho\eta)\cdot\xi\eta\Big)\Big]  
\Big[\rho\xi\Big(\partial_x({1\over\rho\xi})\Big)^2 +\partial_x^2 (-{1\over\rho\xi})\Big] \nonumber\\
&\ & \hspace{0cm}+8{dh\over(\rho\eta)^3}\Big(\partial(\rho\eta)\cdot\partial(\xi\eta)\Big)^2  \nonumber\\
&\ &\hspace{-0.4cm}\quad -4({1\over(\rho\xi)^2})\Big[\Big(\partial(\rho\eta)\cdot d(\xi\eta)\Big)+
\Big(d(\rho\eta)\cdot \partial(\xi\eta)\Big)\Big]\Big(\partial(\rho\eta)\cdot \partial(\xi\eta)\Big)
 \nonumber\\
&\ & \hspace{0cm} 
-{2\over (\rho\xi)^2}\Big[\Big(\rho\eta\cdot d(\xi\eta)\Big)-\Big(d(\rho\eta)\cdot\xi\eta\Big)\Big]\Big[\Big(\partial_x(\rho\eta)\cdot
\partial_x^2(\xi\eta)\Big)-\Big(\partial_x^2(\rho\eta)\cdot\partial_x(\xi\eta)\Big)\Big]
\nonumber\\
&\ & \hspace{0cm}+{8\over (\rho\xi)^3}\Big[\Big(\rho\eta\cdot d(\xi\eta)\Big)-\Big(d(\rho\eta)\cdot\xi\eta\Big)\Big] \Big(\partial_x(\rho\eta)\cdot\partial_x(\xi\eta)\Big)^2\Big\},
\label{gammaexplicit}
\end{eqnarray}
with $(\mu\cdot\nu)\equiv \mu_a\nu^a$  the inner product of SU(2) doublets. 
Here $\rho\xi$ is constrained by
\begin{eqnarray}
\rho\xi=\partial_x h+ \Big(\rho\eta\cdot\partial_x(\xi\eta)\Big)-\Big(\partial_x(\rho\eta)\cdot\xi\eta\Big).
\label{constraint}
\end{eqnarray}
Consequently  $\gamma$ is a function of $h,\rho\eta_a,\xi\eta^a$, which are the lowest component  of the superfields $f,\varphi_a,\varphi^a$ respectively. With this $\gamma$ put in (\ref{KKexplicit})  
 the  $N=(4,0)$ super-Liouville  action (\ref{Omega'}) gets a local expression in 1+1 dimensions as 
\begin{eqnarray}
I= -c\int_{ O_{(0,c)}}\int dxd^4\theta\ y{\cal S}(f,\varphi;x,\theta,t) -{c\over 2}\int_{O_{(0,c)}}\gamma.
 \label{actionlocal}  
\end{eqnarray}
It is important to discuss  non-supersymmetric limit of the theory. 
The purely bosonic parts of $y$ and $\cal S$ are given by\cite{note}
\begin{eqnarray}
{\cal S}(f,\varphi;x,\theta,t)&=&\log\rho\xi+{1\over 2}(\theta\cdot\theta)^2\Big[
-{\partial_x^2\xi\over\xi}-{\partial_x^2\rho\over\rho}+6{\partial_x\xi\over\xi}{\partial_x\rho\over\rho}
\Big]+ O(\eta), \nonumber\\
 y&=&{dh\over \rho\xi}+(\theta\cdot\theta)({d\xi\over\xi}-{d\rho\over\rho}) 
 + {1\over 2}(\theta\cdot\theta)^2
\Big( d\Big[h\partial_x^2(-{1\over\rho\xi})\Big] 
-\partial_x\Big[ hd\partial_x(-{1\over\rho\xi})\Big]\Big)+ O(\eta). \nonumber
\end{eqnarray}
Put them into (\ref{actionlocal}) as well as  $\gamma$ given in (\ref{gammaexplicit}). Calculate the $y\cal S$  as 
$[y]_{\theta^0}[{\cal S}]_{\theta^4}+[y]_{\theta^4}[{\cal S}]_{\theta^0}$. In the action (\ref{actionlocal}) 
 the purely bosonic part of $[y]_{\theta^4}[{\cal S}]_{\theta^0}$ is canceled by the one of ${1\over 2}\gamma$ owing to the constraint  (\ref{constraint}). Then we get   
\begin{eqnarray}
 I= -{c\over 2}\int_{ O_{(0,c)}}\int dx{dh\over \partial_x h}\Big[-{\partial_x^3h\over\partial_x h}+2({\partial_x^2 h\over\partial_x h})^2 + O(\eta) \Big],
 \nonumber
\end{eqnarray}
by using again (\ref{constraint}) as well as $\partial \rho/
\rho=\partial\xi/\xi$.\footnote{ The latter is also a constraint given by (A.1) in \cite{AH2}.}
This is the ordinary non-supersymmetric Liouville theory.  

Thus we have assured ourselves that our arguments are going on a right track.  
However to study symmetries of the theory further it is not convenient to go with the local form of   the action (\ref{actionlocal}). Instead we prefer the non-local form  without using (\ref{gamma})
\begin{eqnarray}
 I= -c \int_{ O_{(0,c)}}\int dxd^4\theta\ y{\cal S}(f,\varphi;x,\theta)  
 -{c\over 2} \int_{\cal M}\int dxd^4\theta\ y\partial_x y,    \label{non-local action}
\end{eqnarray}
in which  $ O_{(0,c)}=\partial{\cal M }$.  
With this we   check the energy-momentum tensor to be given by the $N=(4,0)$ Schwarzian derivative. To this end we need the formula (\ref{superdiff}) for $\delta_v{\cal S}$  and also  the following ones  
\begin{eqnarray}
\delta_v y&=&  dv+
v\partial_x y-y\partial_x v +{1\over 2}D_{\theta c}v D^{\ c}_\theta y +{1\over 2} D^{\ c}_\theta v D_{\theta c} y,  \label{deltay}\\
\delta_v\int dxd^4\theta\ y\partial_x y&=& d\int dxd^4\theta\ 2v\partial_x y,    \label{deltayy}
\end{eqnarray}
which are shown in Appendix B. 
By using these formulae   we find that
\begin{eqnarray}
\delta_v I= -c\int_{ O_{(0,c)}}\int dxd^4\theta\ dv{\cal S}(f,\varphi;x,\theta).
\nonumber
\end{eqnarray}
It may be written in the form
\begin{eqnarray}
\delta_v I= c\int_{O_{(0,c)}}dt \int dxd^4\theta\ v{d\over d t}{\cal S}(f,\varphi;x,\theta).
\nonumber
\end{eqnarray}
Thus the Schwarzian derivative is the energy-momentum tensor of the theory in the left-moving sector. 
When $dv/dt=0$ it is conserved. This was the hallmark of the Liouville theory for the lower
supersymmetric cases.

\section{ PSU(1,1$|$2) symmetry}
\setcounter{equation}{0}

We show that the action (\ref{non-local action}) has symmetry under PSU(1,1$|$2). 
Following \cite{AH2}  such a symmetry 
is non-linearly realized  on a supercoset space PSU(1,1$|$2)/\{SU(2)$\times$U(1)\}
for which the generators of PSU(1,1$|$2) are decomposed as
\begin{eqnarray}
\{T^A\}=\{\underbrace{L,F_a,F^a, \overline L, \overline F_a,\overline F^a}_{{{\rm PSU}(1,1|2)\over{\rm SU}(2)\otimes{\rm U}(1)}},\hspace{-0.2cm}\underbrace{L^0,R^a_{\ b}}_{{\rm SU}(2)\otimes{\rm U}(1)}\hspace{-0.2cm}\}.   \nonumber
\end{eqnarray}
The local coordinates are  the  
 so far discussed superdiffeomorphism $f,\varphi_a,\varphi^a$   and their complex conjugates  $\overline f,\overline\varphi_a,\overline\varphi^a$. They correspond to 
the coset generators $L, F_a,F^a$ and $\overline L, \overline F_a,\overline F^a$. 
 The fermionic coordinates $\varphi_a$ and $\varphi^a$ are doublets of the subgroup SU(2). 
We can calculate the Killing vectors on this coset space following the general method developed in \cite{Ao}. They were worked out in \cite{Ho} and   
 given by\footnote{Precisely speaking  the coset space PSU(2$|$2)/\{SU(2)$\times$U(1)\}  was studied in (\cite{Ho}). The result was adapted for PSU(1,1$|$2)/\{SU(2)$\times$U(1)\} in \cite{AH2}.} 
\begin{eqnarray}
\delta_\epsilon f &\equiv&-i\epsilon_A R^A     \nonumber \\
  &=& \epsilon_L+f\epsilon_{L^0}+(\varphi_c\epsilon_F^{\  c}+\varphi^c\epsilon_{F c})
+\Big(f^2\epsilon_{\overline L}+f(\varphi_c\epsilon_{\overline F}^{\  c}+\varphi^c\epsilon_{\overline  F c})\Big)  \nonumber\\
&\ & +(\varphi_b\varphi^b)(\varphi_c\epsilon_{\overline F}^{\ c}-\varphi^c\epsilon_{\overline Fc})
+(\varphi_c\varphi^c)^2\epsilon_{\overline L},  \label{Killing2a}  \\
\delta_\epsilon \varphi_{a} &\equiv&-i\epsilon_A R^A_{\  a} \nonumber \\
 &=& \epsilon_{Fa}+f \epsilon_{\overline Fa}+{1\over 2}\varphi_a\epsilon_{L^0}-\varphi_c \epsilon_{R\ a}^{\ c}\nonumber \\ 
&\ & +\Big(f\varphi_a\epsilon_{\overline L}+(\varphi_c\varphi^c\epsilon_{\overline Fa}
+2\varphi_c\epsilon_{\overline F}^{\  c}\varphi_a)\Big)
+\varphi_c\varphi^c\varphi_a\varepsilon_{\overline L},       \label{Killing2b} \\
\delta_\epsilon \varphi^{a} &\equiv&-i\epsilon_A R^{A a}  \nonumber \\
 &=& \epsilon_{F}^{\ a}+f \epsilon_{\overline F}^{\ a}+{1\over 2}\varphi^a\epsilon_{L^0}
+\varphi^c\epsilon^{\ a}_{R \ c}\nonumber \\ 
&\ & +\Big(f\varphi^a\epsilon_{\overline L}- (\varphi_c\varphi^c\epsilon_{\overline F}^{\ a}
-2\varphi^c\epsilon_{\overline Fc}\varphi^a)\Big)
-\varphi_c\varphi^c\varphi^a\varepsilon_{\overline L}.   \label{Killing2c}
\end{eqnarray}
Here $\epsilon_A$ are infinitesimal transformation parameters of  PSU(1,1$|$2).  
In \cite{AH2} they have been shown to satisfy the chirality conditions  as well as the superconformal conditions. We then found a remarkable transformation law as
\begin{eqnarray}
\delta_\epsilon \Delta &=&\Big(\epsilon_{L^0}
+2f\epsilon_{\overline L}+2 \varphi_c\epsilon_{\overline F}^{\ c}+2 \varphi^c\epsilon_{\overline F c}\Big)
\Delta, \nonumber
\end{eqnarray} 
for $\Delta$ given by (\ref{Delta}).\footnote{This is the same as (5.8) in \cite{AH2}.} As $d\epsilon_A=0$ and $\partial_x\epsilon_A=0$, 
the quantity $\delta_\epsilon(df+\varphi_a d\varphi^a+\varphi^a d\varphi_a)$ obeys the same transformation law as $\Delta$. Consequently 
 the 1-form $y$ is invariant under the transformations  (\ref{Killing2a})$\sim$(\ref{Killing2c}). 
Using these transformation properties we find the action transform as
\begin{eqnarray}
\delta_\epsilon I= -c\int_{O_{(0,c)}}\int dxd^4\theta\ 
y\Big(\epsilon_{L^0}
+2f\epsilon_{\overline L}+2 \varphi_c\epsilon_{\overline F}^{\ c}+2 \varphi^c\epsilon_{\overline F c}
\Big).  \label{PSU}
\end{eqnarray}
But here we remember that the purely bosonic part of the action is identical with that of the non-supersymmetric Liouville theory. The latter  is invariant under SU(1,1)($\cong$SL(2)), which is a subgroup of PSU(1,1$|$2). It does not admit  such a transformation as  (\ref{PSU}) at all. This observation suggests us that 
 the integration  of (\ref{PSU}) is vanishing. It can be hardly seen in the supercovariant form as it is. So we  have expanded the integrand  in components to examine this. Here also we were involved in 
 massive calculations.
 Finally we have found that the quantities $y,yf,y\varphi_c,y\varphi^c$ are all of the form $\partial_x(\cdots)$.  The details have been  reported in \cite{note}. Thus the $N=(4,0)$ theory has turned out to be invariant under the  symmetry group PSU(1,1$|$2).

\section{Conclusions}
  
In this letter we have formulated the $N=(4,0)$ super-Liouville theory by the coadjoint orbit method
 and have shown that it has all the properties which are characteristic in the lower symmetric Liouville theory, except for one. Namely the  symmetry under PSU(1,1$|$2) has been shown only as a global symmetry above, while for the lower supersymmetric cases it was also a local one in the right-moving sector.  But when promoted the constant parameters $\epsilon^A$ as $d\epsilon\ne 0, \partial_x\epsilon=0$ 
 it is extremely  hard to check the symmetry with either the  non-local form of the action (\ref{non-local action}) or 
  the local one (\ref{actionlocal}). The author hopes to be able to answer the question in a future work. 

To conclude this letter we would like to comment on a prospect for a farther development of the work. 
PSU(1,1$|$2), being a subgroup of PSU(2,2$|$4), is the key symmetry group for the string/QCD duality which was intensively studied 
in the last decade\cite{Ma}. On the string side it is a symmetry for spin-chains, while on the QCD side it is  a symmetry for   
  the  $D=4,\ N=4$ YM  supermultiplets\cite{Bei}. 
In \cite{AH} it was proposed to interpret the spin-chain system by a non-linear $\sigma$-model on a coset space 
 realizing PSU(2$|$2) as a subgroup. But it is worth revisiting the issue of the  string/QCD duality 
 by means of the $N=(4,0)$ super-Liouville theory, since the PSU(1,1$|$2) symmetry has the root of the string theory. The study is in progress.

\renewcommand{\theequation}{\thesection.\arabic{equation}}

\appendix
\section{The lower supersymmetric  Liouville theories}
\setcounter{equation}{0}

\noindent
$N=(0,0)$: \cite{Al}
\begin{itemize}

\item Action: 
\begin{eqnarray}
 I= \int dx\ y\Big[{\cal S}-{1\over 2} \Big({\partial_x^2 f\over \partial_x f}\Big)^2\Big],
\nonumber 
\end{eqnarray}
with 
$$
y={df \over \partial_x f}=dt{\partial_t f \over \partial_x f},\quad\quad\quad 
 {\cal S}= {\partial_x^3 f\over \partial_x f}-
{3\over 2} \Big({\partial_x^2 f\over \partial_x f}\Big)^2.
$$

\item Conformal transformation:
\begin{eqnarray}
\delta_v I&=&-2\int dxdt\ v \partial_t{\cal S}, \nonumber\\
\delta_v y&=& dv+[v\partial_x- \partial_x v]y,\quad\quad 
\delta_v{\cal S}=[v\partial_x +2\partial_x v ]{\cal S} +\partial_x^3v, \nonumber
\end{eqnarray}
with $v=\delta_v x$.

\end{itemize}
$N=(1,0)$:\cite{A2}
\begin{itemize}
\item  $\{D_\theta,D_\theta\}=2\partial_x $.

\item
Constraint:
$
D_\theta f=\varphi D_\theta\varphi.
$
\item Action:
\begin{eqnarray}
I=\int dxd\theta\ y\Big[{\cal S}-{D_\theta^3\varphi\over D_\theta\varphi}{D_\theta^2\varphi\over D_\theta\varphi}\Big]=-2\int dxd\theta\ d\varphi{D_\theta^3\varphi\over (D_\theta\varphi)^2},
\nonumber
\end{eqnarray}
with 
$$
y={df+\varphi d\varphi\over (D_\theta\varphi)^2}=dt{\partial_t f+\varphi \partial_t \varphi\over (D_\theta\varphi)^2},\quad\quad\quad
 {\cal S}= {D_\theta^4\varphi\over D_\theta\varphi}-2{D_\theta^3\varphi\over D_\theta\varphi}{D_\theta^2\varphi\over D_\theta\varphi}.
$$ 

\item Superconformal transformation:
\begin{eqnarray}
\delta_v I&=&-2 \int dx d\theta dt\ v \partial_t{\cal S},\nonumber\\
\delta_v y &=& dv+[v\partial_x  +{1\over 2}D_\theta v D_\theta  -\partial_x v] y, \nonumber\\ 
\delta_v{\cal S}&=&[v\partial_x+{1\over 2}D_\theta v D_\theta + {3\over 2}\partial_x v] {\cal S}+
{1\over 2}D_\theta\partial_x^2 v,   \nonumber
\end{eqnarray}
 with $v=\delta_v  x+\theta\delta_v\theta$.

\end{itemize} 
$N=(2,0)$: \cite{Nieu}
\begin{itemize}

\item $\{D_{\theta +},D_{\theta -}\}=2\partial_x,\quad \{D_{\theta \pm},D_{\theta \pm}\}=0$.

\item 
Constraints: 
$
D_{\theta \pm}\varphi^\mp=0,\quad D_{\theta +}f=\varphi^- D_{\theta +}\varphi^+,\quad  D_{\theta -}f=\varphi^+ D_{\theta -}\varphi^-
$

\item Action: 
\begin{eqnarray}
I=\int dxd^2\theta\ y\Big[{\cal S}+2{\partial_x\varphi^+\over D_{\theta +}\varphi^+}{\partial_x\varphi^-\over D_{\theta -}\varphi^-}\Big]  
= 2\int dxd^2\theta\ d(\log  D_{\theta +}\varphi^+) \log D_{\theta -}\varphi^-, \quad\quad \nonumber
\end{eqnarray}
with 
\begin{eqnarray}
y&=&{df+\varphi^+d\varphi^-+\varphi^-d\varphi^+\over D_{\theta +}\varphi^+D_{\theta -}\varphi^-}
=dt{\partial_t f+\varphi^+\partial_t\varphi^-+\varphi^-\partial_t\varphi^+\over D_{\theta +}\varphi^+D_{\theta -}\varphi^-}, \nonumber\\
 {\cal S}&=& \partial_x(\log D_{\theta +}\varphi^+-\log D_{\theta -}\varphi^-)+2{\partial_x\varphi^+\over D_{\theta +}\varphi^+}{\partial_x\varphi^-\over D_{\theta -}\varphi^-}. \nonumber
\end{eqnarray}

\item Superconformal transformation:
\begin{eqnarray}
\delta_v I&=& -2\int dx d^2\theta dt\ v \partial_t{\cal S},\nonumber\\
\delta_v y &=& dv+[v\partial_x+{1\over 2}D_{\theta +}v D_{\theta -}+{1\over 2}D_{\theta -}v D_{\theta +}
 -\partial_xv]y, \nonumber\\
\delta_v{\cal S}&=& [v\partial_x+{1\over 2}D_{\theta +}v D_{\theta -}+{1\over 2}D_{\theta -}v D_{\theta +}+\partial_xv]{\cal S}+{1\over 2}\partial_x[D_{\theta +}, D_{\theta -}]v, \nonumber
\end{eqnarray}
 with $v=\delta_v x+\theta^+\delta_v\theta^-+\theta^-\delta_v\theta^+$.  
\end{itemize}
Here $\cal S$ is the Schwarzian derivative for the relevant type of the supersymmetry.

\section{Proofs of some formulae}
\setcounter{equation}{0}

We shall prove (\ref{dS}),(\ref{dy}), (\ref{deltay}) and (\ref{deltayy}). 
 The first two formulae 
 may be shown by taking the exterior derivative directly as has been done in Appendix \cite{AH2}, while the last two by the superconformal transformations (\ref{superconf2}) and (\ref{superconf3}). Here we prefer to do it by using the language of the differential geometry. Namely the requirement (\ref{dY}) is equivalent to saying that 
\begin{eqnarray}
i_vdf=\delta_v f,\quad i_vd\varphi_a=\delta_v\varphi_a,\quad  i_vd\varphi^a=\delta_v\varphi^a.     \label{d}
\end{eqnarray}
Here $i_v$  is the known operation of the differential geometry, called anti-derivative,
 while  the superconformal transformation $\delta_v$  
  is known as  the Lie-derivative. Applying (\ref{d}) for $y$ 
gives 
\begin{eqnarray}
i_v y=v.     \label{ivy}
\end{eqnarray}
To find the energy-momentum tensor we  have intentionally assumed $dv\ne 0$. Otherwise  we may set $dv=0$ as usual.  $(y,0)$ in (\ref{yy}) is an element of the  superconformal algebra $\frak g$ 
 such as $(v,l)$ appearing in (\ref{commut}). $\delta_v y$ should be  the superconformal transformation given by (\ref{h,0})
with weight $(-1,0)$. Thus we obtain (\ref{deltay}).  $\delta_vy$ may be alternatively obtained 
\begin{eqnarray}
\delta_vy=dv +i_v dy,   \label{I}
\end{eqnarray}
 by  the identity of the differential geometry
\begin{eqnarray}
\delta_v=di_v +i_vd.    \label{geoidentity}
\end{eqnarray}
Comparing (\ref{I}) with (\ref{deltay}) 
 we know that $dy$ in (\ref{I}) takes the form   (\ref{dy}) owing to  (\ref{ivy}). Finally (\ref{dS}) and (\ref{deltayy}) can be also shown by means of the identity (\ref{geoidentity}). Namely we have  
\begin{eqnarray}
i_vd{\cal S}=\delta_v{\cal S}, \quad\quad di_v \int dx d^4\theta\ y\partial_xy= \delta_v\int dx d^4\theta\ y\partial_xy, \label{F}
\end{eqnarray}
 with 
\begin{eqnarray}
i_v{\cal S}=0, \quad\quad  d \int dx d^4\theta\ y\partial_xy=0.  \nonumber
\end{eqnarray}
By applying (\ref{ivy}) in the l.h.s.s. of the equations in (\ref{F}) it is confirmed that   (\ref{dS}) and (\ref{deltayy}) are right.

\end{document}